\documentstyle[11pt,paspconf,epsfig]{article}

\begin{document}

\title{Cygnus X-1: A Spinning Black Hole? }

\author{Wei Cui\altaffilmark{1}, Wan Chen\altaffilmark{2,3}, and S. N. Zhang\altaffilmark{4,5}}


\altaffiltext{1}{Center for Space Research, Massachusetts Institute of 
Technology, Cambridge, MA 02139, USA} 
\altaffiltext{2}{NASA/Goddard Space Flight Center, Code 661, Greenbelt,
 MD 20771, USA}
\altaffiltext{3}{also Department of Astronomy, University of Maryland, College
 Park, MD 20742, USA} 
\altaffiltext{4}{ES-84, NASA/Marshall Space Flight Center, Huntsville,
AL 35812}
\altaffiltext{5}{also Universities Space Research Association}



\begin{abstract}
It has been a long-standing puzzle in black hole studies what triggers the 
spectral transition between the hard and soft states observed of Cygnus X-1,
a prototypical black hole candidate. In this paper, we will make an attempt 
to address this issue in light of our recent work on black hole spin and 
its observational consequences. 
\end{abstract}


\keywords{black hole physics,black hole spin,accretion disk,Cyg X-1}


\section{Introduction}

Cygnus X-1 is the first dynamically determined black hole (Webster \& Murdin 
1972; Bolton 1972). It is in a binary system with a massive O9.7 Iab 
supergiant, and the orbital period was determined
optically to be 5.6 days. Cyg X-1 is thus intrinsically different from 
the majority of known black holes (BHs) whose companion stars are much less 
massive ($<M_{\odot}$). Curiously, {\it all} high-mass black hole 
binaries (BHBs), including Cyg~X-1, LMC~X-1 and LMC~X-3, are persistent 
X-ray sources, while their low-mass counterparts are exclusively transients. 

Since its discovery, Cyg~X-1 has been considered as an archetypical 
stellar-mass black hole. Its observed spectral and temporal X--ray properties 
have, therefore, often been used to distinguish BHBs from their neutron star 
counterparts. Though seriously flawed, this approach has resulted in the 
discovery of many BHBs whose candidacy was later confirmed by dynamical mass 
measurement based on optical observations. Despite such success, little is 
known about the physics behind many observed phenomena for Cyg~X--1, such as 
its transitions between the hard ($\equiv$ low) and soft ($\equiv$ high) 
states. 

Cyg X-1 spends most of the time in the hard state where the soft X-ray 
(usually 2--10 keV) luminosity is low and the energy spectrum is hard. 
Occasionally (roughly once every 4 years on average), the soft flux suddenly 
jumps up by about a factor of 2 and
the spectrum becomes much softer, i.e., the soft state. Such a state typically
lasts only for several months before the source restores to the normal hard 
state. A complete episode of the hard-to-soft transition, soft state, and 
soft-to-hard transition occurred in 1996. It was discovered by the All-Sky 
Monitor (ASM) on RXTE (Cui 1996) and extensively monitored by both ASM and 
BATSE (Cui et al. 1996; Zhang et al. 1996a,b; also see Cui et al. 1997a,b,c,d,
and Zhang et al. 1997a for detailed studies).

The observed X-ray emission is the result of gravitational energy release 
during mass accretion process. For Cyg X-1, the mass accretion is thought
to be mediated by a so-called ``focused wind'' from the companion star that
nearly fills its Roche-lobe (Gies \& Bolton 1986), unlike transient BHBs
where companion stars overfills the Roche-lobe. The observed orbital 
modulation at both X-ray and radio wavelengths (Zhang, Robinson, \& Cui 1997; 
Fender, Brocksopp, \& Pooley 1997) provides tentative evidence for such
process, because it is thought to be caused by varying 
absorption optical depth through the stellar wind. This interpretation is 
based on the coincidence between the minimum of folded light curves and the 
superior junction (when the companion star is in front of the black hole). 
However, it is still not clear why X-ray modulation {\it only} seems to occur 
in the hard state, or at least it has so far not been observed for the soft 
state. In this paper, we make a critical assumption (but a very probable one) 
that Cyg X-1 is indeed a wind accretion binary. 

\section{Observations and Results}
Figure 1 shows a long-term ASM light curve of Cyg X-1 that covers the entire
1996 episode. The soft state is indicated by a region between the two 
dashed-lines. Besides the rare occurrence of the soft state, the source also 
displays frequent brief flares in the hard state. The flaring activity appears
to be quasi-periodic in nature with a period in the range 3--6 weeks. The 
duration of flares seems to vary from less than one day to tens of
days. The most prominent flare occurred just prior to the hard-to-soft 
transition, and might actually be a characteristic precursor to such an event.
Detailed studies of flaring phenomenon have not been possible, due to the lack 
of high-quality data, so its origin is still unknown. The situation is now 
much improved with the monitoring capability of two complementary all-sky 
monitors, ASM and BATSE. Combining data from both instruments, Zhang et al. 
(1997b) showed that the precursory flare seems to possess some of the spectral
characteristics of the soft state, but the evidence is still tentative. 
\begin{figure}
\psfig{figure=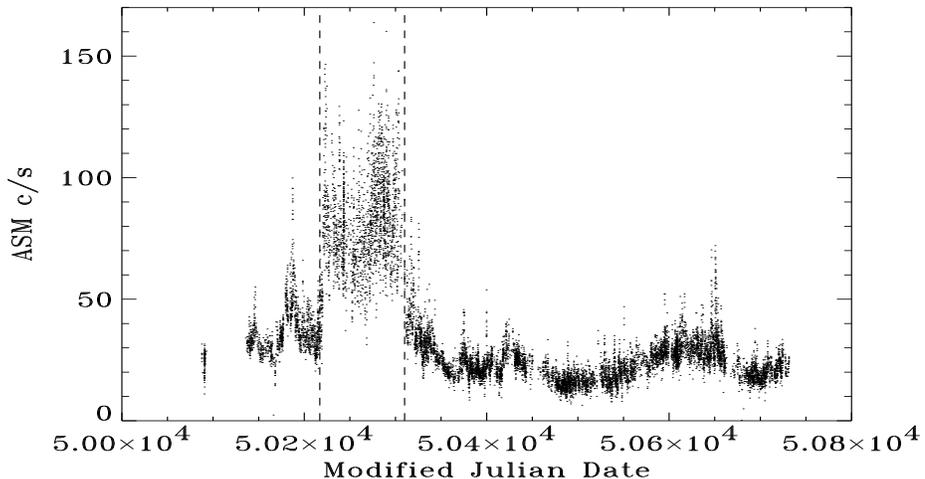,width=5in,height=2.8in}
\caption{The long-term ASM light curve of Cyg X-1. The region between two
dashed-lines indicates the period of soft state. Each data point represents
a measurement with an integration time of 90 seconds. The energy band covered
is between 1.3 and 12 keV. }
\end{figure}

During the transition, the bolometric X-ray luminosity increases only mildly 
(less than a factor of 2; Zhang et al. 1997a), very different from transient 
BHBs where it typically varies by more than one order of magnitude during
a transition. Therefore, a sudden increase of mass accretion rate, due to
some instability, is unlikely responsible for triggering state transitions 
in Cyg X-1. Previously we (Zhang, Cui, \& Chen 1997) noted that the observed 
X-ray spectrum of Cyg X-1 has a distinct ultra-soft component at low energies 
in both the hard and soft states. This spectral component is likely due to the
emission from the innermost region of an optically thick, geometrically 
thin accretion disk. Unlike transient systems where the accretion process is 
probably through advection-dominated flows in the quiescent state (e.g., 
Narayan \& Yi 1994), Cyg X-1 is a persistently luminous BHB. In this system, 
the disk is likely to extend to the last stable orbit, which is determined by 
the mass and spin of the black hole. Given the BH mass of 10-20 $M_{\odot}$ 
for Cyg X-1 from the optical radial velocity measurements (e.g., Gies \& 
Bolton 
1986; Herrero et al. 1995), the BH spin can therefore be derived by measuring 
the location of the inner disk edge. This was achieved by carefully 
characterizing the ultra-soft component (Zhang, Cui, \& Chen 1997). This 
component can be modeled by a multicolor blackbody with the color temperature 
at the inner disk edge only $\sim$0.18 keV, much lower than typical values for 
transient BHBs, and a luminosity $5\times 10^{36}\mbox{ }erg/s$ in the 
hard state. Approaching the soft state, the color temperature increases by a 
factor of $\sim$2.4, and the flux by a factor of $\sim$3. Together they imply 
a decrease of the inner disk radius by a factor of $\sim$3.4, which is {\it 
quantitatively} consistent with a simple reversal of the accretion disk, from 
retrograde to prograde, if the central black hole spins at about 75\% of the 
maximal rate (see Zhang, Cui, \& Chen 1997 for more details). The disk 
reversal in Cyg X-1 is actually a revitalized idea, originally proposed by 
Shapiro and Lightman (1976).

\section{Discussion}
One obvious question is whether disk reversal can actually occur in a real 
binary system. For most BHBs where mass accretion is due to Roche-lobe
overflow of the companion star, the angular momentum of accreting matter
is almost entirely determined by binary orbital motion, thus seems nearly
impossible to imagine a reversal of the accretion flow. For wind accretion
systems like Cyg X-1, however, the fluctuation in residual angular momentum
of the flow behind the hole can cause flip-flop of the disk. This phenomenon
has been demonstrated by both
2D (Matsuda, Inoue, \& Sawada 1987; Benensohn, Lamb, \& Taam 1997) and 3D
(Ruffert 1994a,b, 1995, 1996, 1997) numerical simulations, although the
effects are much suppressed in 3D cases. 
Moreover, Benensohn et al. (1997) showed that the disk flip-flop occurs
quasi-periodically on binary dynamical time scales. {\it Could this be the
origin of brief flares observed in the hard state for Cyg X-1?} Perhaps, the 
soft state is simply a strong, prolonged flare which only occurs on rare 
occasions.
If so, similar spectral evolution would be expected during a flare, just like 
during a state transition. A definitive answer will await detailed, systematic 
studies of the flaring activity in Cyg X-1. 
Unfortunately, the flares are usually very brief, thus are difficult to catch.
A monitoring program with RXTE seems to be the only way in near future to 
obtain essential data for such investigations. 


\acknowledgments

We wish to thank Ron Taam for helpful remarks and discussions during the 
conference.

%
%

%

\end{document}